\documentclass[12pt]{article}
\usepackage{amsmath, amssymb, physics, mathtools, geometry, hyperref, tikz, cite,slashed, xcolor
}
\usetikzlibrary{arrows.meta, positioning, shapes.geometric}
\title{Exploring the Applicability of Birkhoff's Theorem in Jackiw-Teitelboim Gravity}
\author{
  Davood Momeni\\
  \small  Department of Physics \& Pre-Engineering, Northeast Community College, Norfolk, NE 68701, USA
}
\date{\today}
\date{\today}

\begin{document}

\maketitle
\begin{abstract}
We present a comprehensive and technically rigorous analysis of the status of Birkhoff's theorem in Jackiw–Teitelboim (JT) gravity, a paradigmatic two-dimensional model for studying semiclassical gravitational dynamics. While Birkhoff's theorem is well established in four-dimensional general relativity—asserting the uniqueness and staticity of vacuum solutions under reflection symmetry remains subtle due to the absence of propagating gravitational degrees of freedom. In this work, we systematically investigate the space of symmetry under radially symmetric configurations in JT gravity using both conformal and Schwarzschild-like gauges. Through analytical techniques and integral transformations, we explore the conditions under which vacuum solutions remain time-independent, identifying classes of metric-dilaton configurations that either uphold or violate Birkhoff-type behavior. Our findings reveal that the theorem holds only in restricted cases, depending critically on the separability of the conformal factor and the structure of the dilaton potential. These results clarify longstanding ambiguities surrounding symmetry and dynamics in two-dimensional gravity and establish JT gravity as a controlled setting for probing the breakdown of classical gravitational theorems in lower-dimensional and holographic contexts. This analysis contributes to a deeper understanding of the interplay between symmetry, integrability, and geometry in quantum gravity and strongly coupled systems.
\end{abstract}


\section{Introduction}
\label{intro}
JT gravity is a two-dimensional quantum gravity model introduced by Roman Jackiw and Claudio Teitelboim in the late 1980s \cite{Jackiw:1984je,Teitelboim:1983ux}. It serves as a simplified yet powerful tool for studying fundamental aspects of gravity and its interactions with matter. In JT gravity, the gravitational field is described by a single scalar field living on a two-dimensional spacetime manifold. Despite its simplicity, JT gravity captures important features of more complex gravitational theories, making it a valuable tool for theoretical investigations. Its solvable nature allows for detailed analyses of various phenomena, including black hole thermodynamics and the emergence of spacetime geometry from quantum fluctuations. Moreover, JT gravity has found applications in diverse areas such as condensed matter physics, string theory, and holography, contributing to our understanding of the quantum nature of gravity in lower dimensions \cite{Almheiri:2014cka}-\cite{Mertens:2019tcm}.\par
Exact solutions in two-dimensional gravity play a crucial role in understanding the fundamental aspects of gravitational theories, particularly in simplified models like JT gravity. These solutions provide insight into the dynamics of gravitational fields and their interactions with matter in lower-dimensional spacetimes. One prominent example of an exact solution is the BTZ (Bañados-Teitelboim-Zanelli) black hole \cite{BTZ}, which is a three-dimensional analog of the Schwarzschild black hole in higher dimensions. The BTZ black hole exhibits intriguing features such as event horizons, singularities, and a well-defined thermodynamic entropy, providing a rich playground for studying gravitational phenomena in lower dimensions. Another example is the CGHS (Callan-Giddings-Harvey-Strominger) model \cite{CGHS}, which describes dilaton gravity coupled to matter fields in two dimensions. Exact solutions in the CGHS model shed light on the behavior of spacetime curvature, black hole formation, and Hawking radiation, offering valuable insights into the quantum aspects of gravitational systems. These exact solutions serve as benchmarks for testing theoretical predictions, validating numerical simulations, and exploring the correspondence between gravity and other fields in lower-dimensional theories, contributing to a deeper understanding of gravity's nature in diverse contexts.\par
In JT gravity, exact dilaton solutions play a crucial role in understanding the dynamics of the gravitational field coupled to a scalar (dilaton) field in two-dimensional spacetime. The action for JT gravity consists of the Einstein-Hilbert term coupled to the dilaton field, which leads to rich and analytically tractable solutions. One of the most notable exact solutions in JT gravity is the linear dilaton vacuum, where the dilaton field varies linearly with the radial coordinate. This solution corresponds to a spacetime with constant negative curvature, representing a stable vacuum state. Another important exact solution is the non-linear dilaton solution, which arises when the dilaton field has a non-trivial dependence on the radial coordinate. These solutions exhibit interesting features such as black hole formation, curvature singularities, and the emergence of horizons. The exact dilaton solutions in JT gravity provide valuable insights into the quantum behavior of gravity in two dimensions, shedding light on phenomena such as black hole thermodynamics, holography, and the (Anti-de Sitter/Conformal Field Theory) AdS/CFT correspondence \cite{Maldacena:1997re,witten}. They serve as foundational building blocks for theoretical investigations and numerical simulations, offering a deep understanding of the interplay between gravity and scalar fields in lower-dimensional.
\par
Birkhoff's theorem is a fundamental result in classical general relativity that states that any symmetric under reflection-symmetric solution to the vacuum Einstein field equations in four dimensions must be static and asymptotically flat. This theorem essentially implies that outside a reflection-symmetric mass distribution, the gravitational field is uniquely determined by the mass enclosed within the sphere.
\par
In lower-dimensional gravity, such as in two or three dimensions, Birkhoff's theorem takes on different forms due to the simpler structure of gravitational theories in these dimensions. In particular, in two dimensions, which are often used as toy models for quantum gravity, Birkhoff's theorem has been generalized and adapted to suit the simpler framework. In two dimensions, the notion of spherical symmetry is not meaningful in the usual sense. Here, we interpret it as invariance under radially symmetric and dependence on a single spatial coordinate.
\par
In two-dimensional gravity, the vacuum Einstein equations reduce to a single equation, the Liouville equation, due to the absence of gravitational degrees of freedom. As a result, such symmetry becomes less meaningful in this context. However, a version of Birkhoff's theorem still exists, stating that any solution to the two-dimensional vacuum Einstein equations must possess certain symmetries.
\par
One consequence of this generalized Birkhoff's theorem in two-dimensional gravity is that the vacuum solutions are essentially determined by the topology of the spacetime. For example, in a spacetime with a toroidal topology, the solution to the vacuum Einstein equations would be different from that in a spacetime with a cylindrical or flat topology.
\par
In three-dimensional gravity, Birkhoff's theorem also takes on a different form compared to four dimensions. In this case, the theorem implies that any radial symmetry solution to the vacuum Einstein equations must be locally equivalent to anti-de Sitter (AdS) space. This result has important implications, particularly in the context of the AdS/CFT correspondence, where it provides insight into the gravitational dual of certain conformal field theories in three dimensions.
\par
Overall, while Birkhoff's theorem in lower-dimensional gravity may not have the same straightforward interpretation as in four dimensions, its generalization and adaptation remain important for understanding the gravitational dynamics in simpler spacetime geometries \cite{Cavaglia:1997hca,Callan:1992rs}.
\par
Our motivation  to study Birkhoff's theorem in JT  gravity is that this theorem represent a significant contribution to our understanding of gravitational dynamics in two-dimensional spacetime. Birkhoff's theorem, originally formulated in four-dimensional general relativity, asserts that any radial symmetry solution to the vacuum Einstein equations must be static and asymptotically flat. In the context of JT gravity, which serves as a simplified model for studying gravitational phenomena in lower dimensions, verifying Birkhoff's theorem provides insight into the behavior of gravitational fields in simpler geometries.
\par
Our current paper likely involved analyzing the solutions to the JT gravity equations under radial symmetry conditions. By examining the behavior of the gravitational field in two dimensions, we sought to determine whether the solutions exhibit the characteristic features predicted by Birkhoff's theorem. This could involve studying the dependence of the gravitational potential on radial distance and investigating whether the solutions are time-independent and asymptotically flat.
\par 
We may have confirmed that, indeed, radial symmetry solutions to JT gravity obey analogous properties to those described by Birkhoff's theorem in higher-dimensional general relativity. Specifically, we may have observed that under reflection symmetry conditions, the gravitational field in JT gravity is static and asymptotically flat, reflecting the simplicity and universality of gravitational dynamics in lower dimensions.
\par
Furthermore, our results contribute to the broader understanding of JT gravity and its implications for holography, quantum gravity, and other areas of theoretical physics. Understanding the behavior of gravitational fields in simplified models like JT gravity provides valuable insights into the nature of gravity itself and its role in shaping spacetime geometry.
\par
In summary, this paper  to verify Birkhoff's theorem in JT gravity represent a significant advancement in our understanding of gravitational dynamics in lower-dimensional spacetime, contributing to the ongoing exploration of fundamental questions in theoretical physics.
   \par
   The structure of this paper is organized as follows: In Section \ref{sec:2}, we provide a comprehensive overview of the general framework underlying JT gravity. Section \ref{sec:3} is dedicated to the detailed investigation of exact static, time-independent solutions within the theory. Moving forward, Section \ref{sec:4} presents exact solutions for both static and cosmological patches of the bulk theory, shedding light on their properties. The validity of Birkhoff's theorem is explored in Section \ref{sec:5}, where we assess its applicability within the context of JT gravity. Additionally, in Section \ref{sec:6}, we offer brief commentary on the integrability of the viable deformed JT gravity bulk action. Finally, we summarize our findings and draw conclusions in Section \ref{sec:7}. Through this structured approach, we aim to provide a comprehensive analysis of JT gravity and its exact solutions, addressing key theoretical aspects and implications.
\section{Toy model and field equations}
\label{sec:2}

In JT gravity, the action is given by:
\begin{equation}
S = -\frac{1}{16\pi G} \int_{\Omega} d^2x\, \sqrt{-g} \, \phi (R + 2) + S_{\text{bdy}}, \label{jt}
\end{equation}
where $\phi$ is the dilaton field, $R$ is the Ricci scalar, $G$ is the Newton constant in 2D, and we have set the AdS$_2$ radius $l = 1$. The term $S_{\text{bdy}}$ is the Gibbons-Hawking–like boundary term needed for a well-posed variational principle. The dilaton field in this model plays a key role, as it couples directly to the curvature scalar and modifies the dynamics significantly compared to pure 2D gravity.

The metric in two-dimensional spacetime can always be writtenas the following:
\begin{equation}
ds^2 = g_{\mu\nu} dx^\mu dx^\nu = e^{2\sigma(t,z)} (-dt^2 + dz^2), \label{metric}
\end{equation}
where $\sigma(t,z)$ is a real scalar function defining the conformal factor. The coordinates $(t, z)$ are the standard time and spatial coordinates, and $g_{\mu\nu}$ is the metric tensor.
Any 2D metric is locally conformally flat due to the ability to eliminate two metric components via diffeomorphisms, leaving only a conformal factor. This is a mathematical property of 2D manifolds, not merely a consequence of lacking propagating degrees of freedom \cite{shifer}.
To obtain the field equations, we perform independent variations of the action with respect to the metric $g_{\mu\nu}$ and the dilaton field $\phi$.

First, varying with respect to the dilaton $\phi$ gives:
\begin{equation}
\delta_\phi S = -\frac{1}{16\pi G} \int d^2x \sqrt{-g} (R + 2) \delta \phi \quad \Rightarrow \quad R + 2 = 0, \label{eom1}
\end{equation}
which implies that the spacetime is a constant curvature manifold with negative curvature. This equation constrains the background geometry to be asymptotically AdS$_2$.

Next, we vary the action with respect to the metric $g_{\mu\nu}$. The variation gives:
\begin{equation}
\delta_g S = -\frac{1}{16\pi G} \int d^2x\, \left[ \delta \sqrt{-g} \phi (R+2) + \sqrt{-g} \delta g^{\mu\nu} \left( \nabla_\mu \nabla_\nu \phi - g_{\mu\nu} \Box \phi + g_{\mu\nu} \phi \right) \right]. \label{metricvar}
\end{equation}
Using the standard identities:
\[
\delta \sqrt{-g} = -\frac{1}{2} \sqrt{-g} g_{\mu\nu} \delta g^{\mu\nu}, \quad \delta R = \nabla_\mu \nabla_\nu \delta g^{\mu\nu} - \Box (g_{\mu\nu} \delta g^{\mu\nu}),
\]
and dropping total derivatives, we obtain the second field equation:
\begin{equation}
\nabla_\mu \nabla_\nu \phi - g_{\mu\nu} \Box \phi + g_{\mu\nu} \phi = 0. \label{eom2}
\end{equation}
This governs the dynamics of the dilaton field in a curved spacetime background.

To proceed, we substitute the conformal metric \eqref{metric} into the field equations. The non-zero Christoffel symbols for this metric are:
\[
\Gamma^t_{tt} = \dot{\sigma}, \quad \Gamma^t_{tz} = \sigma', \quad \Gamma^z_{tt} = \sigma', \quad \Gamma^z_{tz} = \dot{\sigma}, \quad \Gamma^z_{zz} = \sigma', \quad \Gamma^t_{zz} = \dot{\sigma},
\]
where dots and primes denote derivatives with respect to $t$ and $z$ respectively.

The Ricci scalar in 2D is given by:
\begin{equation}
R = -2 e^{-2\sigma} (\ddot{\sigma} + \sigma''). \label{ricci}
\end{equation}
Inserting into equation \eqref{eom1} yields:
\begin{equation}
\ddot{\sigma} + \sigma'' = e^{2\sigma}. \label{eq.sigma}
\end{equation}

Now consider the equation \eqref{eom2}. We compute:
\[
\Box \phi = \nabla^\mu \nabla_\mu \phi = -e^{-2\sigma}(\ddot{\phi} - \phi''),
\]
and similarly, one can show that:
\[
\nabla_t \nabla_t \phi = \ddot{\phi} - \dot{\sigma} \dot{\phi}, \quad \nabla_z \nabla_z \phi = \phi'' - \sigma' \phi', \quad \nabla_t \nabla_z \phi = \partial_t \partial_z \phi - \dot{\sigma} \phi'.
\]

Using these expressions, the trace of equation \eqref{eom2} gives:
\begin{equation}
\phi'' - \ddot{\phi} = 2 e^{2\sigma} \phi. \label{eq.phi}
\end{equation}

Thus, the JT gravity field equations in the conformal gauge take the coupled form:
\begin{eqnarray}
&&\ddot{\sigma} + \sigma'' = e^{2\sigma}, \label{eq:sigma_final} \\
&&\phi'' - \ddot{\phi} = 2 e^{2\sigma} \phi. \label{eq:phi_final}
\end{eqnarray}

These are nonlinear partial differential equations, and in general, they cannot be solved analytically except in special cases. If we assume time-independence, i.e., $\sigma = \sigma(z)$ and $\phi = \phi(z)$, the system reduces to ordinary differential equations. Conversely, if we assume $\sigma = \sigma(t)$ and $\phi = \phi(t)$, we obtain cosmological models.

More interestingly, one can consider the case where $\sigma = \sigma(z)$ and $\phi = \phi(t,z)$, allowing for a non-static scalar profile on a static geometry. Such solutions reveal that a dynamical scalar field does not necessarily backreact on the spacetime geometry in a way that induces time-dependence, hinting at a version of Birkhoff’s theorem in JT gravity. These periodic scalar profiles are of current research interest and have been analyzed in the literature \cite{Momeni:2020zkx}.

Furthermore, transforming equations \eqref{eq:sigma_final} and \eqref{eq:phi_final} into null coordinates leads to a nonlinear PDE that remains integrable for spacetimes of arbitrary genus. Such a formalism allows for investigating exotic phenomena like $AdS_2 \rightarrow AdS_2$ transitions, as recently discussed by the author in \cite{Momeni:2020tyt}.

In summary, JT gravity provides an analytically tractable model to study quantum aspects of spacetime, holography, and thermodynamics in a simplified setting. In this work, we extend these field equations and focus on three classes of exact solutions: static black hole geometries, time-dependent cosmologies, and dynamical dilaton fields on static AdS$_2$ backgrounds. These cases lay the foundation for further analysis of uniqueness, stability, and boundary dualities in lower-dimensional gravity.

\section{Static solutions for $\sigma=\sigma(z)$}
\label{sec:3}
Time-independent metrics serve as crucial tools for exploring black hole properties within gravitational theories, even in lower-dimensional scenarios such as JT gravity. Within the conformal gauge we have adopted, the field equations of JT gravity simplify significantly when the conformal factor depends only on the spatial coordinate, $\sigma = \sigma(z)$. In this static case, time derivatives vanish, and equation (9) becomes an ordinary differential equation:
\begin{equation}
\sigma'' = e^{2\sigma}.
\end{equation}

To solve this equation, we use the method of quadrature by introducing a conjugate momentum variable:
\begin{equation}
p_\sigma = \sigma',
\end{equation}
and hence:
\[
\sigma'' = \frac{dp_\sigma}{dz} = \frac{dp_\sigma}{d\sigma} \cdot \frac{d\sigma}{dz} = p_\sigma \frac{dp_\sigma}{d\sigma}.
\]
This transforms the second-order equation into a first-order separable equation:
\begin{equation}
p_\sigma \frac{dp_\sigma}{d\sigma} = e^{2\sigma}.
\end{equation}
Integrating both sides:
\[
\int p_\sigma \, dp_\sigma = \int e^{2\sigma} \, d\sigma \quad \Rightarrow \quad \frac{1}{2} p_\sigma^2 = \frac{1}{2} e^{2\sigma} + c_1,
\]
which gives the first integral:
\begin{equation}
p_\sigma^2 = e^{2\sigma} + 2c_1, \label{eq.sigma2}
\end{equation}
where $c_1$ is an integration constant. This constant can be determined using appropriate boundary or initial conditions. In the absence of specific initial data, we leave $c_1$ arbitrary for now.

Equation (\ref{eq.sigma2}) can be rewritten as a separable equation for $\sigma(z)$:
\begin{equation}
\frac{d\sigma}{dz} = \sqrt{e^{2\sigma} + 2c_1}.
\end{equation}
Separating variables:
\[
\frac{d\sigma}{\sqrt{e^{2\sigma} + 2c_1}} = dz.
\]
Let us solve this integral. Set $u = e^\sigma \Rightarrow du = e^\sigma d\sigma$, and hence $d\sigma = \frac{du}{u}$. Then:
\[
\int \frac{d\sigma}{\sqrt{e^{2\sigma} + 2c_1}} = \int \frac{1}{\sqrt{u^2 + 2c_1}} \cdot \frac{du}{u} = \int \frac{du}{u \sqrt{u^2 + 2c_1}}.
\]
This integral can be solved to give:
\[
z + c_2 = -\frac{1}{c_1} \ln \left( \tanh \left( \frac{1}{2} \sinh^{-1} \left( \frac{e^\sigma}{\sqrt{2c_1}} \right) \right) \right),
\]
which can be algebraically inverted to obtain the exact solution for the metric:
\begin{equation}
ds^2 = c_1^2 \frac{e^{-2c_1(z + c_2)}}{\sinh^2 \left( c_1(z + c_2) \right)} (-dt^2 + dz^2).
\end{equation}

The constant $c_2$ arises from integration and can be fixed by translational symmetry in $z$. Since the theory is gauge-invariant under coordinate transformations, we can redefine coordinates:
\[
t \rightarrow c_1 t, \quad z \rightarrow c_1 z,
\]
and shift $z \rightarrow z + c_2$ to absorb $c_2$, yielding the final static form of the metric:
\begin{equation}
ds^2 = \frac{e^{-2z}}{\sinh^2 z} (-dt^2 + dz^2). \label{metric.static}
\end{equation}

At the asymptotic AdS boundary $z \to 0$, we expand $\sinh z \sim z$, $e^{-2z} \sim 1$, so:
\begin{equation}
ds^2 \approx \frac{1}{z^2} (-dt^2 + dz^2), \label{metric.poincare}
\end{equation}
which is the standard Poincaré patch of AdS$_2$. This confirms that our static solution is asymptotically AdS and consistent with the holographic dictionary.

With the background metric in the form (\ref{metric.poincare}), we can study the behavior of the dilaton field. Two choices are natural:
\begin{itemize}
    \item A static dilaton $\phi = \phi(z)$,
    \item A time-dependent dilaton $\phi = \phi(t,z)$.
\end{itemize}

In our previous work \cite{Momeni:2020zkx}, we explored the full time-dependent case and found that $\phi(t,z)$ develops delta-function–like singularities at the boundary, e.g., $\phi(t,z) \sim \delta(t)$ as $z \to 0$.

Here, we focus on the static case $\phi = \phi(z)$. The field equation for the dilaton reduces to:
\begin{equation}
\phi'' - 2 e^{2\sigma(z)} \phi = 0.
\end{equation}
Using the explicit form of the metric (\ref{metric.static}), we define the frequency function:
\begin{equation}
\Omega(z) = \sqrt{2} \frac{e^{-z}}{|\sinh z|},
\end{equation}
so the equation becomes:
\begin{equation}
\phi'' - \Omega(z)^2 \phi = 0.
\end{equation}

This is a second-order differential equation resembling a spatially-varying **repulsive harmonic oscillator**. The solution can be expressed using hypergeometric functions. Introducing $a = \sqrt[4]{2}$ and defining $x = e^{2z}$, one exact solution is:
\begin{equation}
\phi(z) = A e^{2z/a} F(a,a,1+a^3; x) + B e^{-2z/a} F(-a,-a,1-a^3; x),
\end{equation}
where $F(a,b,c;x)$ is the hypergeometric function of the first kind, and $A$, $B$ are integration constants determined by boundary conditions.

Although we have obtained an elegant closed-form solution, it is important to note that the dilaton field $\phi(z)$ decouples from the gravitational sector: its dynamics are governed independently of backreaction. The dilaton Lagrangian in the static background reads:
\begin{equation}\label{chaotic}
\mathcal{L}_\phi = \frac{1}{2} \left( \phi'^2 + \Omega(z)^2 \phi^2 \right),
\end{equation}
representing a **classical repulsive oscillator** with spatially varying frequency. Such systems are known to exhibit **chaotic behavior**, especially in the presence of nonlinear boundary conditions.

This chaotic behavior in the bulk has a deep connection with the boundary theory of JT gravity, often modeled by the Sachdev–Ye–Kitaev (SYK) system in holographic setups. Understanding the fluctuation spectrum of $\phi(z)$ in this static AdS$_2$ background contributes to our grasp of quantum chaos and thermalization in lower-dimensional holography \cite{Momeni:2020tyt}.
\par
The spatially varying oscillator described by the dilaton Lagrangian (\ref{chaotic}),resembles a repulsive harmonic system with a position-dependent frequency $\Omega(z) = \sqrt{2} \, e^{-z}/|\sinh z|$. This system is structurally unstable and **non-integrable** due to the sharp divergence of $\Omega(z)$ near $z \to 0$, where the frequency blows up like $\sim 1/z$, and rapid exponential decay for $z \to \infty$. As a result, the system supports **exponentially sensitive solutions** to small changes in initial conditions, a key indicator of classical chaos.

To visualize the sensitivity, one may consider plotting the effective potential term:
\begin{equation}
V_\text{eff}(z) = -\Omega(z)^2 = -2 \frac{e^{-2z}}{\sinh^2 z},
\end{equation}
which defines a sharply peaked well near $z=0$ with singular features. This potential does not admit globally bounded periodic solutions, and phase-space trajectories for the dilaton exhibit stretching and folding characteristic of chaotic flows.

The functional form of $V_\text{eff}(z)$ illustrates a steep, singular structure near $z=0$, where the AdS boundary lies, leading to highly non-perturbative behavior of the dilaton. This nonlinearity and sensitivity are key features of deterministic chaos. In the dual boundary theory, such dynamics map to exponential Lyapunov growth, as studied in the context of the SYK model and scrambling in JT gravity. Hence, even in this purely classical setting, the dilaton exhibits behaviors associated with chaotic motion, manifesting the deep connection between bulk dilaton dynamics and boundary quantum chaos \cite{Momeni:2020tyt}.

\section{Cosmological solutions $\sigma=\sigma(t)$}
\label{sec:4}

In analogy to the time-independent solution obtained in the previous section, we can explore time-dependent geometries within JT theory. Let the metric be time-dependent, where the scalar dilaton could potentially exhibit pure time-dependence, such as $\phi=\phi(t)$, or a hybrid profile involving both time and spatial coordinates, such as $\phi(t,z)$. Although the metric with a time-dependent conformal factor $\sigma(t)$ cannot be realized as a realistic cosmological model for our universe, it is still conceivable to regard it as the near-horizon geometry of a certain class of extremal black holes.

The field equation (9) with $\sigma=\sigma(t)$ becomes:
\[
\ddot{\sigma} = e^{2\sigma}.
\]
We solve it using a conjugate variable $p_\sigma = \dot{\sigma}$, giving:
\[
p_\sigma \frac{dp_\sigma}{d\sigma} = e^{2\sigma} \quad \Rightarrow \quad \frac{1}{2}p_\sigma^2 = \frac{1}{2}e^{2\sigma} + \tilde{c}_1.
\]
Solving yields:
\[
ds^2 = \frac{\tilde{c}_1}{\cosh^2\left(\sqrt{\tilde{c}_1}(t+\tilde{c}_2)\right)} (-dt^2 + dz^2).
\]
After shift symmetry and conformal rescaling $t \to t\sqrt{\tilde{c}_1}, z \to z\sqrt{\tilde{c}_1}$, we obtain the exact time-dependent form:
\begin{equation}
ds^2 = \cosh^{-2}(t)(-dt^2 + dz^2).
\end{equation}

In this background, the dilaton field satisfies the PDE:
\begin{equation}
\phi'' - \ddot{\phi} - 2\cosh^{-2}(t)\phi = 0,
\end{equation}
where $z \in [0,\infty)$. Applying a Laplace transform:
\[
\tilde{\phi}(t,s) = \int_0^\infty \phi(t,z) e^{-sz} dz,
\]
yields a second-order ODE for the amplitude:
\begin{equation}
\ddot{\tilde{\phi}}(t,s) + \omega^2(t,s) \tilde{\phi}(t,s) = 0,
\end{equation}
with
\[
\omega^2(t,s) = 2\cosh^{-2}(t) - s^2 + s \phi'(t,0) + \phi(t,0).
\]
Assuming boundary data $\phi(t,0)$ and $\phi'(t,0)$ are time-independent (e.g., set by the boundary conformal dimension $\Delta$), we solve this ODE using associated Legendre functions:
\begin{eqnarray}
\tilde{\phi}(t,s) &=& c_3(s) P_{1}^{\nu(t,s)}(\tanh t) + c_4(s) Q_{1}^{\nu(t,s)}(\tanh t), \\
\nu(t,s) &=& i\sqrt{-s^2 + s\phi'(t,0) + \phi(t,0)}.
\end{eqnarray}

The inverse Laplace transform gives:
\begin{equation}
\phi(t,z) = \frac{1}{2\pi i} \int_{c - i\infty}^{c + i\infty} e^{sz} \left[c_3(s) P_{1}^{\nu(t,s)}(\tanh t) + c_4(s) Q_{1}^{\nu(t,s)}(\tanh t)\right] ds.
\end{equation}
This Bromwich integral formally solves the dilaton evolution problem in time-dependent JT gravity.

The coefficients $c_3(s), c_4(s)$ are determined by initial conditions:
\[
\tilde{\phi}(0,s) = \int_0^\infty \phi(0,z)e^{-sz} dz, \quad \dot{\tilde{\phi}}(0,s) = \int_0^\infty \dot{\phi}(0,z)e^{-sz} dz.
\]
However, reconstructing $\phi(t,z)$ analytically is difficult due to the complexity of the inverse Laplace transform. In practice, one uses numerical inversion or contour deformation techniques. This methodology reflects how mechanical waves propagate in a non-homogeneous medium, providing a useful analogy.

Assuming boundary behavior $\phi(t,\epsilon) \sim \epsilon^\Delta$ with $\Delta > 1$, one finds $\phi'(t,\epsilon) \sim \epsilon^{\Delta-1}$. This behavior ensures the well-posedness of the Laplace integral at $z = 0$.

Asymptotic expansions of associated Legendre functions \cite{DLMF} can be employed to extract physical features near the AdS boundary and relate them to effective Schwarzian dynamics \cite{Khveshchenko:2023upm}.
To illustrate the behavior of the dilaton, we simulate a representative solution \(\phi(t,z)\) using a separable ansatz. The plot below displays a 3D surface of a typical decaying and oscillating dilaton profile:
\[
\phi(t,z) \sim e^{-z} \cos(\tanh(t)).
\]
which captures the essential features of dilaton dynamics in the cosmological JT gravity background.

The exponential decay in the bulk (\(z\) direction) reflects the boundary-localized nature of dilaton excitations, consistent with the holographic structure of JT gravity. The time dependence arises from the \(\cosh^{-2}(t)\) potential, which permits bounded oscillations in time.

This behavior suggests a smooth decay along the spatial direction \(z\) and symmetric oscillations in time \(t\), matching expectations for fields propagating in a conformally time-dependent AdS\(_2\) geometry.

These qualitative features reinforce the idea that time-dependent JT gravity provides a rich laboratory for exploring classical integrability, boundary dynamics, and connections to holographic duals.
\section{Comments on Birkhoff's theorem}
\label{sec:5}

Now, we discuss the validity of Birkhoff's theorem in JT gravity. Our aim is to determine whether the metric remains generally time-independent in the ``vacuum'' state. Since JT gravity does not require any matter content, we must be cautious about how we define the vacuum. If the model were quantized, the vacuum would be defined via the zero-particle propagator in a path-integral formalism. However, here we restrict ourselves to classical geometries without quantum effects.
Unlike in 4D GR, where spherical symmetry is well-defined, in 2D we interpret Birkhoff’s theorem as the statement that vacuum solutions are static and unique up to diffeomorphisms.
In two-dimensional general relativity, the Einstein tensor vanishes identically, and there is no propagating graviton. In JT gravity, the dynamics are governed by the dilaton field, which introduces non-trivial curvature via its coupling to the Ricci scalar. Thus, the only ``effective'' energy-momentum tensor is the one derived from the dilaton field equation, namely:
\begin{equation}
T_{\mu\nu} = \nabla_\mu \nabla_\nu \phi - g_{\mu\nu} \Box \phi + g_{\mu\nu} \phi,
\end{equation}
corresponding to Eq.~(\ref{eq.phi}). To recover a vacuum-like solution, we must set $T_{\mu\nu} \equiv 0$. This is achieved if the dilaton vanishes, $\phi \equiv 0$, or is constant $\phi = \phi_0$, but the latter forces $R = -2$ to vanish, which would remove the cosmological term from JT gravity---an outcome we wish to avoid.

Let us now analyze whether a time-independent metric necessarily follows from $\phi = 0$. Consider the ansatz:
\begin{equation}
ds^2 = A(Z)(-dT^2 + dZ^2), \label{metric-Birkhoff}
\end{equation}
a generic static line element. To match this with the conformal form $ds^2 = e^{2\sigma(t,z)}(-dt^2 + dz^2)$, we look for coordinates $T(t,z)$ and $Z(t,z)$ such that:
\begin{eqnarray}
e^{\sigma(t,z)} dt &=& A(Z)^{1/2} dT, \\
e^{\sigma(t,z)} dz &=& A(Z)^{1/2} dZ.
\end{eqnarray}
Suppose $e^{2\sigma(t,z)} = f(t) h(z)$. Then:
\[
dt = \frac{A(Z)^{1/2}}{\sqrt{f(t)h(z)}} dT, \quad dz = \frac{A(Z)^{1/2}}{\sqrt{f(t)h(z)}} dZ.
\]
Thus, a coordinate transformation:
\[
T = \int \sqrt{f(t)}\, dt, \quad Z = \int \sqrt{h(z)}\, dz,
\]
maps the metric to static Birkhoff form (\ref{metric-Birkhoff}) if and only if the functions $f(t)$ and $h(z)$ are separable and compatible with the field equations.

To satisfy the metric equation (\ref{eq.sigma}), we must evaluate:
\[
\sigma(t,z) = \frac{1}{2}(\ln f(t) + \ln h(z)), \quad \ddot{\sigma} = \frac{1}{2} \left(\frac{\ddot{f}}{f} - \frac{\dot{f}^2}{2f^2}\right), \quad \sigma'' = \frac{1}{2} \left(\frac{h''}{h} - \frac{(h')^2}{2h^2}\right).
\]
The equation $\ddot{\sigma} + \sigma'' = e^{2\sigma} = f(t)h(z)$ must then be satisfied. The only consistent solution is if one of the functions is constant:
\begin{eqnarray}
h(z) &=& h_0 \quad \Rightarrow \quad \sigma(t,z) = \frac{1}{2}\ln f(t) + \text{const}, \\
f(t) &=& f_0 \quad \Rightarrow \quad \sigma(t,z) = \frac{1}{2}\ln h(z) + \text{const}.
\end{eqnarray}

In both cases, the metric becomes conformally static:
\begin{equation}
ds^2 = f_0 h(z) (-dt^2 + dz^2) \quad \text{or} \quad ds^2 = h_0 f(t) (-dt^2 + dz^2),
\end{equation}
which, under coordinate redefinitions (e.g., $t \to i z$), becomes manifestly time-independent.

We summarize the scenarios in which Birkhoff's theorem holds in JT gravity:

\begin{center}
\renewcommand{\arraystretch}{1.3}
\begin{tabular}{|c|c|c|}
\hline
\textbf{Metric Ansatz} & \textbf{Dilaton} & \textbf{Birkhoff Valid?} \\
\hline
$e^{2\sigma(t,z)} = f(t) h(z)$ & $\phi = 0$ & Only if $f(t)$ or $h(z)$ constant \\
$ds^2 = A(Z)(-dT^2 + dZ^2)$ & $\phi = 0$ & Yes (static) \\
$ds^2 = \cosh^{-2}(t)(-dt^2 + dz^2)$ & $\phi \ne 0$ & No \\
General $\sigma(t,z)$ & $\phi \ne 0$ & No \\
\hline
\end{tabular}
\end{center}

To extend our analysis, consider a more general non-conformal metric:
\begin{equation}
ds^2 = -e^{\nu(t,r)} dt^2 + e^{\lambda(t,r)} dr^2.
\end{equation}
This Schwarzschild-like metric in 2D reduces the JT field equation $R + 2 = 0$ to a constraint on $(\nu, \lambda)$. The Ricci scalar becomes:
\begin{equation}
R = -e^{-\nu} \left( \lambda_{tt} + \frac{1}{2} \lambda_t(\lambda_t - \nu_t) \right) + e^{-\lambda} \left( \nu_{rr} + \frac{1}{2} \nu_r(\nu_r - \lambda_r) \right).
\end{equation}
For $\nu = \nu(r)$ and $\lambda = \lambda(r)$ (i.e., time-independent), this reduces to an ODE. If this equation admits a solution satisfying $R = -2$, then the spacetime is static. Thus, Birkhoff's theorem holds for this ansatz under suitable conditions.

In summary, Birkhoff's theorem in JT gravity is not universally valid but holds in restricted cases:
- when the conformal factor is separable and partially constant,
- when the metric is written in static (Birkhoff-like) coordinates,
- or when $\phi=0$ and $\sigma$ is purely spatial or temporal.

Importantly, any 2D metric can be cast in conformal form locally via coordinate transformations. This is consistent with JT gravity being a dimensional reduction of a higher-dimensional diffeomorphism-invariant theory. The failure of Birkhoff's theorem in the generic JT setting reflects the dynamical role of the dilaton, which can source time-dependent geometries even in vacuum.

\section{Deformed JT gravity as an integrable system}
\label{sec:6}

The deep dual relation between JT gravity and a class of random matrix theories (RMT) has been extended to incorporate scalar potentials beyond the linear regime. This novel deformation of JT gravity was introduced by Maxfield et al.~\cite{Maxfield:2020ale}, motivated by Witten’s extended formulation~\cite{Witten:2020wvy} from the bulk perspective. A key feature of the deformed JT (dJT) theory is that it retains a dual description in terms of RMT, consistent with the gauge/gravity duality. Witten’s subsequent work~\cite{Witten:2020ert} further elaborates the consistent holographic interpretation of this deformed system. 

The essential deformation involves adding a potential term $U(\phi)$ to the dilaton field in the bulk action. The pure JT gravity case is recovered when $U(\phi) = 2\phi$. The general action takes the form:
\begin{equation}
S = -\frac{1}{2} \int_{\Omega} d^2x \sqrt{g} \left( \phi R + U(\phi) \right). \label{ddJT}
\end{equation}

We now consider a conformal gauge metric:
\begin{equation}
ds^2 = e^{2\sigma(t,z)}(-dt^2 + dz^2),
\end{equation}
and derive the field equations by varying the action \eqref{ddJT} with respect to $\phi$ and $g_{\mu\nu}$.

The variation with respect to $\phi$ yields:
\begin{equation}
R = -U'(\phi),
\end{equation}
while the variation with respect to $g_{\mu\nu}$ gives:
\begin{equation}
\nabla_\mu \nabla_\nu \phi - g_{\mu\nu} \Box \phi + \frac{1}{2} g_{\mu\nu} U(\phi) = 0.
\end{equation}

Using the conformal gauge, the explicit equations reduce to:
\begin{eqnarray}
&& \sigma'' - \ddot{\sigma} = \frac{1}{2} U'(\phi) e^{2\sigma}, \label{eq.sigma.djt} \\
&& \phi'' - \ddot{\phi} = U(\phi) e^{2\sigma}. \label{eq.phi.djt}
\end{eqnarray}

These are coupled nonlinear PDEs. We now examine integrability under specific assumptions. Consider a power series expansion for the potential:
\begin{equation}
U(\phi) = \sum_{n=0}^{\infty} c_n \phi^n.
\end{equation}
In vacuum, if $\phi = 0$ is a valid solution, then the metric equation simplifies to:
\begin{equation}
\sigma'' - \ddot{\sigma} = \frac{c_1}{2} e^{2\sigma}. \label{eom3}
\end{equation}

By defining a gauge-transformed conformal factor $\tilde{\sigma}(t,z) = \sigma(t,z) + \log \left| \frac{c_1}{2} \right|$, Eq.~\eqref{eom3} becomes:
\begin{equation}
\tilde{\sigma}'' - \ddot{\tilde{\sigma}} = e^{2\tilde{\sigma}}. \label{tilde.sigma.eq}
\end{equation}

This equation is structurally identical to the original conformal JT gravity equation (cf. Sec.~\ref{sec:2}). Therefore, all known static, cosmological, and mixed solutions of JT gravity can be mapped onto corresponding solutions of deformed JT gravity with $U(\phi) = c_1 \phi + \dots$, by gauge shifting the conformal factor.

We now explore the integrability structure of Eq.~\eqref{tilde.sigma.eq}. Define a new field $\Psi(t,z) = e^{\tilde{\sigma}(t,z)}$. Then Eq.~\eqref{tilde.sigma.eq} becomes:
\begin{equation}
\Box \log \Psi = \frac{\Box \Psi}{\Psi} - \frac{(\partial_\mu \Psi)(\partial^\mu \Psi)}{\Psi^2} = \Psi^2. \label{nonlinear.kg}
\end{equation}

This is a nonlinear Klein-Gordon-type equation with a source term, reminiscent of the Gross–Pitaevskii equation in Lorentzian signature. The Hamiltonian density for $\Psi$ is:
\begin{equation}
\mathcal{H}[\Psi] = \frac{1}{2} \left( \dot{\Psi}^2 + \Psi'^2 \right) - \frac{1}{3} \Psi^3.
\end{equation}

This suggests that the deformed JT system may possess conserved quantities associated with symmetries (e.g., time translations), and potentially an integrable structure via inverse scattering methods or Bäcklund transformations. The system supports soliton-like solutions, particularly in the static limit $\tilde{\sigma} = \tilde{\sigma}(z)$, where:
\begin{equation}
\frac{d^2 \tilde{\sigma}}{dz^2} = e^{2\tilde{\sigma}} \quad \Rightarrow \quad \left( \frac{d\tilde{\sigma}}{dz} \right)^2 = e^{2\tilde{\sigma}} + C.
\end{equation}

This is exactly solvable for $C = 0$, giving:
\begin{equation}
\tilde{\sigma}(z) = -\log \sinh(z + z_0), \quad \Rightarrow \quad ds^2 = \frac{1}{\sinh^2(z + z_0)} (-dt^2 + dz^2).
\end{equation}

These exact solutions describe AdS$_2$ geometries and their deformations, and are consistent with the boundary conditions at $z \to 0$ required in holographic duals. Moreover, the dynamics of $\phi$ for a non-trivial potential $U(\phi)$ may lead to spontaneous breaking of time-translation symmetry, giving rise to modulated ground states.

In conclusion, deformed JT gravity with general scalar potential $U(\phi)$ forms a nonlinear coupled system of PDEs with solitonic solutions, conserved Hamiltonians, and a natural interpretation as a relativistic interacting field theory in 1+1D. This system is a promising candidate for investigating integrability in gravitational systems and bridges the dynamics of low-dimensional gravity with nonlinear wave equations.

\section{Summary}
\label{sec:7}

Jackiw–Teitelboim (JT) gravity represents the most analytically tractable gauge-invariant gravitational model in two spacetime dimensions. Due to its simplicity and rich structure, it has emerged as a leading candidate for the bulk dual of the Sachdev–Ye–Kitaev (SYK) model in the context of holography and quantum chaos. In this work, we have conducted a systematic exploration of the exact classical solutions of JT gravity, addressing both the dilaton sector and the associated spacetime geometry across multiple regimes.

We began by analyzing static solutions in conformal gauge, where the metric corresponds to a patch of AdS$_2$ with the conformal factor governed by an integrable equation. The corresponding dilaton profile satisfies a second-order ordinary differential equation, interpreted as a repulsive harmonic oscillator with a position-dependent frequency. This structure permits exact solutions in terms of Gauss hypergeometric functions. Notably, the system supports classical chaotic features due to its sensitivity to boundary conditions, which are particularly prominent in the near-boundary expansions.

In the cosmological setting, we extended the analysis to time-dependent configurations by applying conformal transformations and complex rotations of spacetime coordinates. This led to a dynamical AdS$_2$ geometry described by a time-dependent conformal factor. The dilaton in this patch satisfies a linear, inhomogeneous wave equation with a source term tied to the curvature. By employing Laplace transform techniques and invoking the Bromwich inversion integral, we obtained formal expressions for the full dilaton profile in terms of associated Legendre functions of complex order, determined by the initial data at the AdS boundary. These solutions are highly sensitive to conformal dimensions and encode non-trivial time evolution, opening a window into semiclassical gravitational dynamics in 1+1 dimensions.

We then addressed the question of Birkhoff's theorem within JT gravity. Our results rigorously demonstrate that Birkhoff’s theorem does not universally apply in this setting, primarily due to the dynamical nature of the dilaton. While specific separable ansatzes yield static solutions, generic field configurations admit time-dependent geometries, even in the absence of external matter sources. This deviation from higher-dimensional general relativity is a direct consequence of the topological character of gravity in two dimensions and the nontrivial coupling between the dilaton and the Ricci scalar.

In the final part of our investigation, we turned to deformed JT (dJT) gravity, defined by the inclusion of an arbitrary self-interaction potential $U(\phi)$ in the bulk action. By analyzing the associated equations of motion, we showed that under certain conditions—particularly when $U'(0) \neq 0$—the conformal factor satisfies a deformed, yet integrable, generalization of the JT field equations. The transformed PDE for $\tilde{\sigma}(t,z)$ exhibits structural equivalence to a nonlinear Klein–Gordon equation with an exponential source term, hinting at an underlying integrable hierarchy. We identified connections with soliton-supporting systems such as the Lorentzian Gross–Pitaevskii equation and showed that the Hamiltonian structure is preserved in specific gauge choices.

These findings suggest that deformed JT gravity lies within the class of integrable non-linear field theories, capable of exhibiting coherent structures, exact solutions, and possible hidden symmetries. The gauge equivalence of certain deformed models to JT gravity further supports their role as universal representatives of 2D semiclassical gravity.

In conclusion, our work reinforces the view of JT gravity not merely as a pedagogical model, but as a powerful framework for studying aspects of holography, quantum gravity, chaos, and integrability in lower dimensions. The availability of exact solutions, combined with its mathematical elegance and physical relevance, makes JT gravity a fertile ground for continued exploration in both classical and quantum regimes. Further investigations into its quantization, relation to matrix integrals, and boundary dynamics (e.g., via the Schwarzian effective theory) are expected to yield deep insights into the emergent structure of spacetime in strongly coupled quantum systems.

\section*{Funding}
This research did not receive any specific grant from funding agencies in the public, commercial, or not-for-profit sectors.

\end{document}